\documentclass[doublecol]{epl2} 

\title{Taxonomy and clustering in collaborative systems: \\ the case of the on-line encyclopedia Wikipedia}
\shorttitle{Taxonomy and clustering in Wikipedia} 

\author{A. Capocci\inst{1} \and F. Rao \inst{2} \and G. Caldarelli\inst{3,2}}
\shortauthor{A. Capocci \etal}

\institute{                    
  \inst{1} Dip. di Informatica e Sistemistica
Universit\`a ``Sapienza'', via Ariosto, 25 00185 Rome, Italy\\
  \inst{2} Centro Studi e Ricerche e Museo della Fisica
``E. Fermi'', Compendio Viminale, 00185-Rome, Italy \\
  \inst{3} SMC Centre, INFM-CNR, Dip. di Fisica, Universit\`a ``Sapienza'',
P.le A. Moro 2, 00185-Rome, Italy
}

\pacs{89.75.Fb}{Structures and organisations in complex systems}
\pacs{89.75.Hc}{Networks and genealogical trees}
\pacs{89.75.-k}{Complex systems}

\abstract{ In this paper we investigate the nature and structure of
the relation between imposed classifications and real clustering in a
particular case of a scale-free network given by the on-line
encyclopedia Wikipedia. We find a statistical similarity in the
distributions of community sizes both by using the top-down approach of
the categories division present in the archive and in the bottom-up
procedure of community detection given by an algorithm based on the
spectral properties of the graph. Regardless the statistically similar
behaviour the two methods provide a rather different division of the
articles, thereby signaling that the nature and presence of power laws
is a general feature for these systems and cannot be used as a
benchmark to evaluate the suitability of a clustering method.  }

\begin{document}

\maketitle

Many real systems can be modeled by means of a scale-free network
\cite{book1,book2}. By such mathematical representation it is often
possible to better understand the development of these systems and
possibly to discover some unexpected behaviour. Much scientific
interest has recently focussed on their community structure, often
revealed by highly clustered regions of a network. Dividing a
network into communities of nodes sharing some given property gives a
coarse grained representation of the whole system. The paramount
example of such is given by information networks such as the World
Wide Web (WWW). The WWW is a network composed by html documents
connected by hyperlinks and for its giant structure only partial
studies on its community structure have been produced
\cite{web1,web2}. Once a large information network such as the WWW is
decomposed into communities, data mining can be performed in a more
efficient way by restricting the data search to smaller regions of the
WWW where the desired information is more probable to be found.

Another well--known example of information network is the on--line,
user--generated encyclopedias Wikipedia available at
http://www.wikipedia.org in several languages. Articles of each
encyclopedia can be represented as nodes, and the hyperlinks from an
article to another within the Wikipedia form directed networks shaped
by the article creations and edits of thousands of individual users
around the world. The versions of Wikipedia we explored display
statistical properties\cite{bro1,zlat,Capo2} typical of complex
networks such as the WWW, whose Wikipedia is a subset, even though
their microscopic growth processes differ noticeably: while in the
first case users need ``administrator'' rights to edit webpages,
Wikipedia articles can be edited by any user.

Wikipedia networks have varying sizes depending on the language and
activity of the underlying users' community and ranging from a few
hundreds to more than one million articles. Here we present an
analysis of a sample of this set of graphs (hereafter Wikigraphs)
collected in September of 2007 from the web site
http://download.wikimedia.org/. In particular, we study similarities
and differences between two possible classifications of Wikipedia
articles: their internal categorization and the partition, by a
suitable algorithm, of the network formed by articles and hyperlinks
between them.

Wikipedia articles are gathered into categories according to their
topics. The classification of articles, the creation or the deletion
of categories are decided upon the agreement of the whole Wikipedia
community. In turn, categories are organised hierarchically according
to their generality. However, articles and categories do not strictly
form a perfect tree, since an article or a category may happen to be
the child of more than one parent category. Therefore, the taxonomy of
articles can be represented a direct acyclic graph
\cite{wiki_categories}.

Much work has been devoted to the study of the statistical features of
taxonomies, in order to understand whether their overall properties
could reveal any general pattern of organization.  In most of the
cases, one observes power-law distributions in the number of
offsprings that can be explained by means of Yule processes or by the
inherent properties of supercritical trees.  The first explanation has
been proposed for taxonomies of natural species\cite{willis} showing a
power-law decay in the frequency of the number of species for a given
genus. Based on such data, Yule\cite{yule} introduced a model to
explain how mutations in a population of individuals may eventually
form a series of different species in the same genus.  The results of
this process have a rather good agreement with the observed data.
Yule processes represent a fundamental mechanism in the production of
power laws, though they do not reproduce completely the richness of
the scale-invariance presented in natural taxonomies.  Indeed, when
looking at the statistical distribution of the sizes of trees (which
corresponds to the distribution of genera in the same family and
different families in the same order) a similar power-law relation has
also been found\cite{Bur,Bur2}. In this case, the value of the
power-law exponent may also depend upon the observed ecosystem type
\cite{cec}.
Following this experimental evidence, one may decide to model the
development of the whole hierarchical tree of the taxonomy by using a
random branching process. It has been analytically shown that the
subtree size distribution of a random tree displays a power-law decay,
$P (s) \propto s^{-\tau} $. The exponent $\tau$ is $3/2$ for critical
random trees, where the branching number is 1 \cite{harris}, and
equal to $2$ \cite{paoletto} if the branching number is larger than
$1$ as it is often the case in several real instances of growing
networks. Therefore, the presence of power laws with an exponent
nearby $2$ can be considered just a consequence of the parent--child
structure of a taxonomy\cite{cec2}. 

Beside their classification in categories, Wikipedia articles may also
be clustered by the analysis of the network that, through hyperlinks,
connect them. Such task is nowadays performed by a number of
algorithms \cite{danon05}. Methods based on edge betweenness and clustering
coefficient assume that edges lying on most of the shortest paths in
the graph or with low clustering coefficient are likely to connect
separate communities. By recursively deleting the edges with larger
betweenness or low clustering, the graph splits into its communities
\cite{GN02,radicchi04}. Methods that optimise the network modularity,
instead, form cluster of nodes so that the density of link within the
communities are maximized against the number of links among communities
\cite{donetti04,clauset04}. Finally, spectral methods are based on the
analysis of the eigenvalues and eigenvectors of suitably chosen
functions of adjacency matrix $A$, whose size is given by the number
of vertices $n$ in a graph and whose elements $a_{ij}$ are equal to
$1$ if an edge exists between nodes $i$ and $j$ and zero otherwise
\cite{donetti04,kleinberg99,Capo}.

While any of such method can be applied in small graphs, unfortunately
they turn to be unusable in larger networks since they require
exceeding computational resources or time. Though, the detection of
strongly interconnected communities of nodes in a network can still be
achieved by finding the attraction basins of random walks on the
graph. This is obtained through the method we adopted in our
investigation, the MCL algorithm, which provides a fast response in a
reasonable time even for networks including thousands of nodes, and
can be tuned opportunely in order to maintain its efficiency for even
larger systems. However, it has to be noted that the MCL algorithm too
is unable to cluster the larger available Wikigraphs.

The MCL algorithm \cite{MCL1,MCL2} finds the partition of a network at
the desired resolution as follows:
\begin{enumerate}
\item 
start with the transition matrix $A$ of the network and normalise 
each column of the matrix to obtain a stochastic matrix $S$; 
\item 
compute $S^{2}$; 
\item 
take the $p^{th}$ power ($p>1$) of every element of $S^{2}$ and
normalise each column to one;
\item 
go back to step $2$. 
\end{enumerate}
The physical meaning of this procedure is the following: through step
$2$ we compute the probability that a random walk visits edges two
steps apart the starting position. If a walk starts within a communities,
with greater probability it will remain inside it. By raising these
probabilities to a power (step $3$) and then normalising them, we
enhance these paths with respect to the others. The effect is to
create a statistical matrix $S'$ corresponding to an adjacency matrix
(and hence a graph) in which edges between communities are removed.

After some iterations, MCL converges to a matrix $S_{MCL(p)}$ which is
invariant under transformations $2$ and $3$.  Only a few lines of
$S_{MCL(p)}$ have non-zero entries, yielding the nodes' clusters as
separated basins (there is in general exactly one non-zero entry per
column).  As noted above, the step $3$ reinforces the high probability
walks at short time scale at the expense of the low probability
ones. The whole process of iteration, on physical grounds, of the MCL
algorithm corresponds to simulating many random walks on the networks
and strengthening their flow where it is already strong and weakening
it where it is weak. The parameter $p$ tunes the granularity of the
clustering. If $p$ is large, the effect of step $3$ becomes stronger
and the random walks are likely to end up in a greater number of
smaller basins of attraction, or communities. On the other hand, a small
$p$ produces larger communities. In the limit of $p=1$, only one cluster
is found.  The MCL method, thus, has a parameter to be tuned,
determining the resolution of the resulting division of the network.
In order to compare the communities emerging from the MCL analysis and
the taxonomy established by Wikipedia contributors, we set such
parameter $p$ to produce approximately the same number of categories
observed in the data.

In our work, we have investigated the relation between the category
structure of Wikipedia and the clustering properties of the underlying
graph representing articles and hyperlinks between them. The category
system is a tool to let users browse the content of Wikipedia with
greater ease. Due to the large amount of information to be handled by
users, a self--organised categorization system would help in
classifying pages without human intervention and
discussion. Clustering methods, often based on the topology of a
graph, are used at this aim, especially to deal with user--generated
content on the WWW \cite{hotho06,voss07,watanabe07}. However, the application
of automatic clustering methods in each specific context has to be
validated by comparing their yielding to the results of manual
indexing. The aim of this paper, thus, is to compare and discuss the
partition of the graph based on the built-in taxonomy, i.e. the
categories, and the one obtained by means of the MCL algorith, applied
to the network of the Wikipedia pages connected by internal links.
The various data sets analyzed can be downloaded from the archive of
the encyclopedia (http://download.mediawiki.org). We selected some
data dumps selected according the number of articles $S$. In
particular, the largest set we considered is the English Wikipedia
(En, $2,042,361$ articles), while the smallest one is the Norman one
(Nrm, $2,750$ articles), as reported in table \ref{wikipedias}. They
span several order of magnitude in the number of nodes, ranging at
that time from the few thousands articles of the {\em Norman} archive
to the millions articles of the {\em English} one. This way, we are
able to check whether finite size effects affect our observations. For
each Wikipedia, we analyzed the datasets reporting the category
structure and the internal link structure.
\begin{table}[h!]
\begin{center}
\begin{tabular}{|l|c|r|} 
\hline              
Language & Wikipedia & Articles \\
\hline
English & En & 2042361 \\
\hline            
German & De & 650241 \\\hline     
Italian & It & 357538 \\\hline            
Norwegian & No & 134943 \\\hline            
Catalan & Ca & 81660  \\\hline
Danish & Da & 70757 \\\hline  
Croatian & Hr & 35932 \\\hline                        
Galician & Gl & 28113 \\\hline                             
Simple English & Simple & 19921 \\\hline
Latin & La & 15602 \\\hline                                       
Neapolitan & Nap & 12603 \\\hline
Occitan & Oc & 10359 \\ \hline
Afrikaan & Af & 8443\\ \hline
Aragonese & An & 7144 \\ \hline
Venetian & Vec & 5974 \\ \hline
Corsican & Co & 5324 \\ \hline
Interlingua & Ia & 3652 \\ \hline
Alemannic & Als & 3141 \\ \hline
Norman & Nrm & 2750 \\ \hline
\end{tabular} 
\caption{The Wikipedia versions sampled in the analysis and their size.}
\label{wikipedias}
\end{center}
\end{table}

\begin{figure}
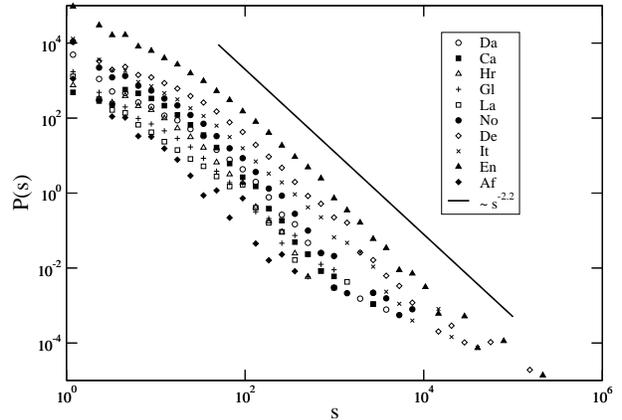

\onefigure[width=8cm]{cat-size-distribution.eps}
\caption{The frequency of category sizes in a sample of the Wikigraphs
analyzed here. The solid line represents $s^{-2.2}$.}
\label{fig2} 
\end{figure}

We start by considering the category and cluster size distributions,
i.e. the distribution of the number of Wikipages contained in a
category or in a cluster. The statistical properties of Wikipedia
taxonomies appear to display a remarkable regularity in the size
distribution of categories. As shown in Fig.\ref{fig2} the category
size distribution $P(s)$ is heavy-tailed, following approximately a
power law $P(s) \propto s^{-\gamma}$ with $\gamma \simeq 2.2$ for very
different sizes of the system.

\begin{figure}
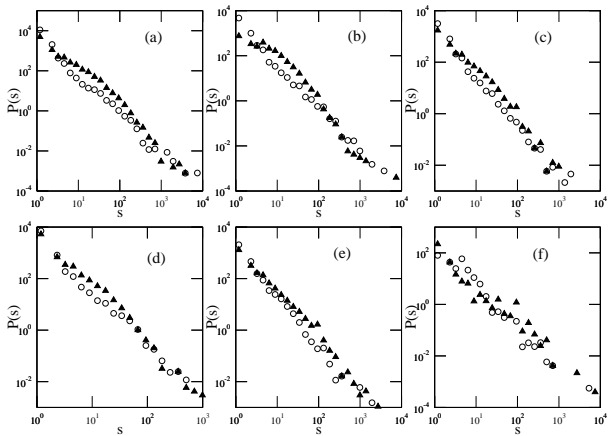

\onefigure[width=8cm]{cat_vs_clus-size-distribution.eps}
\caption{ The category size distribution (triangles) compared to the
cluster size distribution (circles) obtained by the MCL algorithm for
the Danish (a), Croatian (b), Galician (c), Simple English (d), Latin
(e), Neapolitan (f) Wikipedia.}
\label{fig3} 
\end{figure}

We then applied the MCL algorithm to measure the size distribution of
topology-based communities. Unfortunately, our survey has to limit itself
to Wikipedia smaller than a given size, above which the problem of
clustering the network becomes computationally
intractable. Interestingly, the clustering-based partition we obtain
follow a very similar cluster size distribution with a power--law
decay for large values of $s$, as it can be observed in figure
\ref{fig3}. In the experiment, we have tuned the granularity parameter
of the MCL algorithm in order to obtain approximately the same number
of communities and categories.

Nonetheless, Fig.\ref{fig3} shows only a similar partition structure,
which does not necessarily imply that the partitions themselves are
similar. To compare the two partitions, we adopt as a measure the {\em
adjusted Rand index} as it has been recently generalised to soft
partitions \cite{Campello}. Standard Rand index \cite{rand71} results
from a pairwise comparisons of the elements in two different
partitions $P,Q$.  If we denote as
\begin{itemize}
\item[a:] Number of pairs of data objects belonging to the same class
in $P$ and to the same class in $Q$.

\item[b:] Number of pairs of data objects belonging to the same class
in $P$ and to different classes in $Q$.

\item[c:] Number of pairs of data objects belonging to different
classes in $P$ and to the same class in $Q$.

\item[d:] Number of pairs of data objects belonging to different
classes in $P$ and to different classes in $Q$.
\end{itemize}
we can compute the Rand index $R$ as
\begin{equation}
R=\frac{a+d}{a+b+c+d}=\frac{2(a+d)}{n(n-1)}
\end{equation}
since $a+b+c+d$ is given by the total number $n(n-1)/2$ of pairs in
the system.This is a measure of the agreement between partitions since
terms $a$ and $d$ contribute to consistent classifications
(agreements), whereas terms $b$ and $c$ are measures of inconsistent
classifications (disagreements).  Unfortunately, in the case of a
partition composed by many clusters, the $d$ element dominates such
that the quantity $R$ can be close to $1$ even if the partitions
substantially differs.  To overcome this, the adjusted Rand index
$R^a$ has been introduced,
\begin{equation}
R_a=\frac{a-\frac{(a+c)(a+b)}{a+b+c+d}}
{\frac{2a+b+c}{a+b+c+d}-\frac{(a+c)(a+b)}{a+b+c+d} },
\end{equation}
which is equal to $0$ if the two partitions $P$ and $Q$ are randomly
drawn \cite{HA85}. 

It has to be noticed that articles in Wikipedia can lie in more than
one category. Therefore, the taxonomy has to be treated as a soft
partition, i.e. a partition where classes intersection is not null and
elements can belong to more than one class with varying
intensity. Accordingly, we adopted the generalization of the adjusted
Rand index for fuzzy partitions recently introduced \cite{Campello}.

\begin{figure}
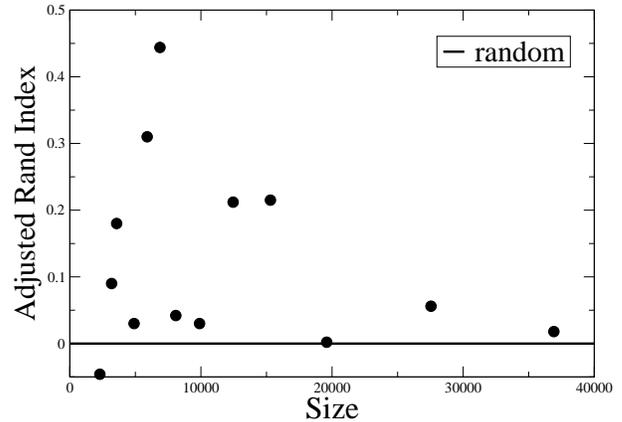

\onefigure[width=8cm]{adj_rand_index.eps}
\caption{The overlap between category-based and clustering-based
partitions measured by the Adjusted Rand Index for a sample of
Wikipedia networks. The red solid line represents the expected value
for two random partitions.}
\label{fig4}
\end{figure}

The Adjusted Rand Index takes very different values when measured in
different systems. Moreover, its value seems to be uncorrelated with
the network size, as reported in figure \ref{fig4} where only articles
assigned to at least one category are taken into account. This shows
that the categorization of Wikipedia articles does not necessarily
correspond to the clustering patterns emerging from the MCL
algorithm. The latter, in fact, could results in some case in a quite
different organization of knowledge. 

From all the above analysis we can conclude that the two divisions of
the graphs represent truly different local and global processes on the
network, depending upon the decentralised users' action and the
consensual collective choice respectively.  This behaviour does not
reflects into a different frequency distribution $P(s)$ of the
category and cluster sizes. Rather, this quantity is distributed with
the same scale-invariant distribution given by $P(s) \propto
s^{-2.2}$.  This suggests that the presence of power-laws in these
quantities is more related to the fractal nature of the branching in
the category structure\cite{paoletto} when approaching the problem
from top to down, or conversely to the Zipf's law \cite{Zipf} when
considering the inverse bottom-up process of cluster formation. The
varying agreement between clustering and categorization across the
studied versions of Wikipedia suggests that links in Wikipedia do not
necessarily imply similarity or relatedness relations. From a
technological point of view, this observation implies that, before
switching to automatic categorization of items in Wikipedia and in
other information networks, it should be tested how the selected
clustering algorithm performs with respect to manual indexing.

\acknowledgments
A.C. and G.C. acknowledge the European project DELIS for support.
Authors acknowledge enlightening discussions with Stefano Leonardi.


\begin{thebibliography}{99}

\bibitem{book1}
  \Name{Caldarelli G.}
  \Book{Scale-Free Networks}
  \Publ{Oxford University Press, Oxford}
  \Year{2007}.
  
\bibitem{book2}
  \Name{Buchanan M.} 
  \Book{Nexus} 
  \Publ{W. W. Norton \& Co., New York} 
  \Year{2002}.


\bibitem{web1} 
\Name{Adamic L.A. \and Huberman B.A.} 
\REVIEW{Science}{287}{2000}{2115}.

\bibitem{web2} 
\Name{Donato D., Millozzi S., Leonardi S.\and Tsaparas P.} 
\REVIEW{Proceedings of the 8$^th$ International Workshop on the Web and Databases}{WebDB}{2005}{}.

\bibitem{bro1} 
\Name{Broder A., Kumar R., Maghoul F., Raghavan P., Rajagopalan S., 
Stata R., Tomkins A. \and Wiener J.} 
\REVIEW{Computer Networks}{33}{2000}{309}.

\bibitem{zlat} 
\Name{Zlati\'c V., Bo\v{z}i\v{c}evi\'c M., \v{S}tefan\v{c}i\'c H. \and Domazet M.}
\REVIEW{Physical Review E}{74}{2006}{016115}. 

\bibitem{Capo2} 
\Name{Capocci A., Servedio V.D.P., Colaiori F., Buriol
L.S., Donato D., Leonardi S. \and Caldarelli G.}  
\REVIEW{Physical Review E}{74}{2006}{036116}.

\bibitem{wiki_categories} For more detailed guidelines in the
categorization of Wikipedia articles, see the web page
http://en.wikipedia.org/wiki/Wikipedia:Categorization

\bibitem{willis}
\Name{Willis J.C. \and Yule G.U.}
\REVIEW{Nature}{109}{1922}{177}.

\bibitem{yule}
\Name{Yule G.U.} 
\REVIEW{Philosophical Transaction of the Royal Society of London, Series B}{213}{1925}{21}

\bibitem{Bur} 
\Name{Burlando B.}
\REVIEW{Journal of Theoretical Biology}{146}{1990}{99}.

\bibitem{Bur2} 
\Name{Burlando B.}
\REVIEW{Journal of Theoretical Biology}{163}{1993}{161}.

\bibitem{cec} 
\Name{Caretta Cartozo C., Garlaschelli D., Ricotta C., Barthelemy M., Caldarelli G.}
q-bio.PE/0612023, preprint 2006.

\bibitem{harris}
\Name{Harris T.E.} 
\Book{The Theory of Branching Processes} 
\Publ{Springer-Verlag, Berlin}
\Year{1963}.

\bibitem{paoletto} 
\Name{De Los Rios P.} 
\REVIEW{Europhysics Letters}{56}{2001}{898}. 

\bibitem{cec2} 
\Name{Caldarelli G., Caretta Cartozo C., De Los Rios P. \and Servedio V.D.P.}
\REVIEW{Physical Review E}{69}{2004}{035101(R)}.

\bibitem{danon05}
\Name{Danon L., D\'{i}az-Guilera A., Duch J \and Arenas A.}
\REVIEW{J. Stat. Mech}{2005}{P09008}.

\bibitem{GN02} 
\Name{Girvan M. \and Newman M.E.J.}
\REVIEW{Proceedings of the National Academy of Science (USA)}{99}{2002}{7821}.

\bibitem{radicchi04}
\Name{Radicchi F., Castellano C., Cecconi F., Loreto V. \and Domenico
  Parisi D.}
\REVIEW{Proceedings of the National Academy of Sciences}{101}{2004}{2658}

\bibitem{donetti04}
\Name{Donetti L. \and Mu\~oz M.-A.}
\REVIEW{J. Stat. Mech.: Theor. Exp.}{}{2004}{10012}

\bibitem{clauset04}
\Name{Clauset A., Newman M.E.J. \and Moore C.}
\REVIEW{Phys. Rev. Lett.}{70}{2004}{066111}

\bibitem{kleinberg99}
\Name{Kleinberg J.M.}
\REVIEW{Journal of ACM}{46}{1999}{604}

\bibitem{Capo} 
\Name{Capocci A.,Servedio V.D.P.,Caldarelli G. \and Colaiori F.} 
\REVIEW{Physica A}{352}{2005}{669}.

\bibitem{hotho06} 
\Name{Hotho A., J\"aschke R., Schmitz C. \and Stumme G.}  
\REVIEW{Proceedings of the Workshop on Applications of Semantic
Technologies}{}{2006}{}

\bibitem{MCL1}
\Name{Enright A.J., Van Dongen  S. \and Ouzounis C.A.}
\REVIEW{Nucleic Acid Research}{30}{2002}{1575}

\bibitem{MCL2}
\Name{Van Dongen S.}
\REVIEW{Ph.D. thesis (University of Utrecht)}{}{2000}

\bibitem{voss07}
\Name{Voss J}
\REVIEW{Proceedings 10th Internat. Symposium for Information Science, Constance}{}{2007}{243}

\bibitem{watanabe07} 
\Name{Watanabe Y., Asahara M.  \and Matsumoto Y.}  
\REVIEW{Proceedings of the 2007 Joint Conference on Empirical
Methods in Natural Language Processing and Computational Natural
Language Learning (EMNLP-CoNLL)}{}{2007}{649}



\bibitem{Campello} 
\Name{R.J.C.B. Campello} 
\REVIEW{Pattern Recognition Letters}{28}{2007}{833-841} 

\bibitem{rand71} 
\Name{W.M. Rand} 
\REVIEW{J. Amer. Statist. Assoc}  {1971} {846-50}.

\bibitem{HA85} 
\Name{L. Hubert \and P. Arabie} 
\REVIEW{J. Classif.} {2} {1985} {193-218} 

\bibitem{Zipf} 
\Name{Zipf G.K.} 
\Book{Human Behavior and the Principle of Least Effort} 
\Publ{Addison-Wesley, Reading MA} 
\Year{1949}.


\end{thebibliography}
\end{document}